\documentclass[11pt,twoside]{article}

\usepackage{asp2006}
\usepackage{epsf}
\usepackage{psfig}
\usepackage{lscape}

\begin{document}
\title{Modelling the formation and evolution of disk galaxies}
\author{M.~J.~Stringer$^{1,2}$ and A.~J.~Benson$^1$}
\affil{$^1$Department of Astrophysics, Keble Rd., Oxford, OX1 3RH, U.K.}
\affil{$^2$Theoretical Astrophysics, Caltech, MC130-33, 1200 E. California
Blvd., Pasadena, CA 91125, U.S.A.}

\begin{abstract} 
Inspired by recent work on feedback in disk galaxies (Efstathiou 2000,
Silk 2003) and on the angular momentum distribution in simulated gas
halos (Sharma and Steinmetz 2005), a fully dynamic model of disk
galaxy formation and evolution has been developed. This is used to
demonstrate how observed galactic systems could have formed from halos
similar to those found in simulations and applies physically motivated
models of star formation and feedback to explore whether the true
nature of these processes would be manifest from local and
cosmological observables. This is made possible by computational
integration with the galaxy formation model developed originally by
the group at Durham University (Cole et al. 2000).
\end{abstract} 

Galaxies within the {\sc Galform} hierarchical model can now be
decomposed into $\sim10$pc regions so that their formation can be
followed without any prior assumption of disk profile. Their
subsequent evolution can be studied across the entire disk, as
described in \citet{Stringer07}.

\begin{figure}
\plotfiddle{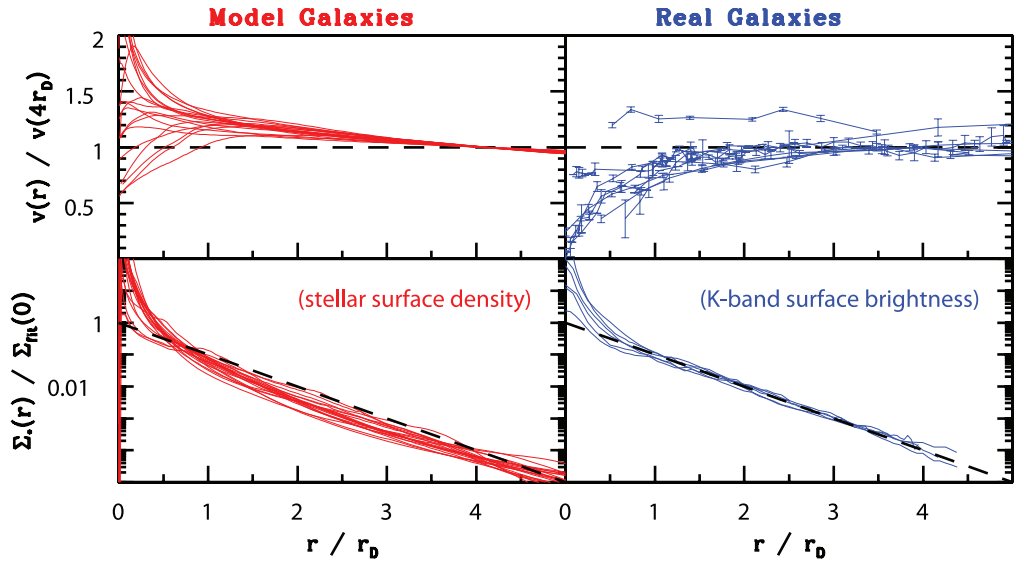}{3in}{0}{100.0}{100.0}{-200}{-180}
%\plotfiddle{Stringer_Fig1_bw.eps}{3in}{0}{100.0}{100.0}{-200}{-180}
\caption{The radial profiles of real and modeled galaxies, each in
radial units of its own particular scalelength, $r_{\rm D}$.  Rotation
curves, in units of their value at four scalelengths, are plotted in
the top row. Stellar surface density and surface brightness are
plotted in the second row in units of the central fitted value. The 11
real galaxies are from \citet{Kassin06}, and have bulge fractions of
20\% or less. The modeled systems are taken from a realisation of
hundreds of hierarchically forming halos, of which the 20 here are
those central galaxies which have the same range of mass and bulge
fraction as their real counterparts.}\label{profiles}
\end{figure}

\begin{figure}
\plotone{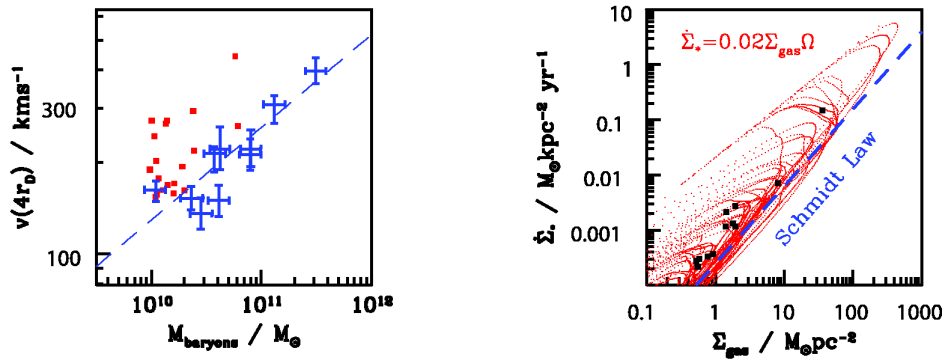}
%\plotone{Stringer_Fig2_bw.eps}
\caption{Global Properties of the galaxies from Figure
\ref{profiles}. {\bf Left:} The circular velocity, at four
scalelengths, for the real systems (error bars, with a dotted line of
best fit) and those from the model (squares). {\bf Right:} Star
formation rates from the modeled galaxies, which applied a ``Silk
law'' of star formation, as given in the equation in the
figure. Values are plotted both for every individual annular region
(dots) and for each disk as a whole (squares). The dotted blue line
shows the traditional observational relationship from
\citet{Kennicutt98}, $\dot{\Sigma}_\star=
2.5\times10^{-4}\left[\Sigma_{\rm gas}/M_\odot{\rm
pc}^{-2}\right]^{1.4}M_\odot{\rm kpc}^{-2}$.}\label{global}
\end{figure}

Three initial findings are presented here. Firstly, the predicted
stellar surface density profiles, governed predominantly by the inital
angular momentum distribution of the halos, are qualitatively similar
to those observed, exhibiting the same key features of an exponential
decrease and a central peak.  Secondly, the predicted circular
velocities are in poor agreement with both observed rotation curves
(Fig.~\ref{profiles}) and the Tully-Fisher relation
(Fig.~\ref{global}).

Finally, the model suggests that a constant {\em local} star formation
efficiency per orbit may produce regions which significantly deviate
from a Schmidt-Kennicutt law whilst the {\em global} star formation
rates could still follow a trend with comparable slope and scatter to
the observations.

\acknowledgements 
MJS acknowledges a PPARC studentship at the University of Oxford and
the hospitality of the CTCP at Caltech. AJB acknowledges support from
the Gordon and Betty Moore Foundation.

\end{document}